\documentclass[12pt]{article}
\usepackage{graphicx}
\usepackage{amssymb,amsmath,amsfonts,palatino,amsthm}
\usepackage{amssymb}
\usepackage{epstopdf}
\usepackage{slashed} 
\newcommand{\f}{\begin{equation}}
\newcommand{\ff}{\end{equation}}

\begin{document}

\title{Natural and bionic neuronal membranes:  possible sites for quantum biology
}
\author{Lee Smolin\thanks{lsmolin@perimeterinstitute.ca} 
\\
\\
Perimeter Institute for Theoretical Physics,\\
31 Caroline Street North, Waterloo, Ontario N2J 2Y5, Canada}
\date{\today}
\maketitle

\begin{abstract}

A new concept for bionic quantum technology is presented based on a hybrid of a silicon wafer on which is layered a phospholipid membrane, such as is found in biological cell membranes.  The phosphorus atoms in the head groups of the membranes carry nuclear spins which serve as memory qubits.  The role of control qubits may be played by unpaired spins of extra electrons on phosphate groups with a single negative charge, in polar,  zwitterionic headgroups such as  phosphatidylcholine (PC).   
Classical control gates and circuits are embedded on the silicon wafer, as in proposals by Kane and others for solid state quantum computing devices.

A proposal to extend these ideas to neuronal membranes, which makes use of the chirality of the phospholipid molecules that make up its bulk, is also briefly sketched.  The chirality of the phospholipid molecules is argued, at least at low temperature, to induce Chern-Simons couplings, which may give rise to robust qubits in topological winding states, defined on the cylinder mod punctures-which are the ion channels.


\end{abstract}

\newpage

\tableofcontents


\section{Introduction}

There is a long history of speculation that quantum mechanics could be important for the functioning of biological systems\cite{wil,Roger}.  These would go beyond the obvious fact that only quantum mechanically bound atoms and molecules have the precise specificity of chemical and electrical properties needed for all biological molecules to function\cite{wil}.  

In a recent and very provocative paper\cite{MF}, Matthew Fisher proposes that the nuclear spins of 31 Phosphorus (31P) are uniquely suited as carriers of quantum information in hypothetical instances in which quantum coherence and entanglement could play a role in biological systems.   This is partly due to the fact that phosphorus is, besides hydrogen, the unique element common in biology which has a spin $\frac{1}{2}$ nucleus.   

However, note that phosphate groups are ubiquitous in biology.  Three key places they are found are the backbones of $DNA$ and $RNA$, in $ATP$ and in the phospholipid molecules that are the primary ingredient of biological membranes.

It would be extremely interesting to extend Fisher's speculations to find roles for quantum coherence in the functioning of these central biological systems. 
One way to do this is presented in the next two sections; the basic idea is to use the phosphorus in biological membranes, or model systems engineered to mimic them, in the construction of quantum devices for quantum information processing.
Then, in section 4 I sketch another route, based on the chirality of neuronal membranes.

Like all new proposals for quantum technology, there are serious technical challenges which must be overcome before either of these proposals can be realized.  The purpose of this brief paper is only to propose the concepts; further exploration would be needed before any claims could be made about their practicality.

In \cite{Kane} Kane proposed an architecture for a quantum computer, since called a Kane, or solid state, quantum computer,  which is to be composed of a silicon wafer doped with phosphorus atoms\cite{rp}.  To describe the new idea proposed here, it will help to first describe the basic set up of a Kane quantum computer.

The nucleus of phosphorus in the most common isotope (31P) has a spin $J=\frac{1}{2}$ and these serve as memory qubits.  The nuclear magnetic dipoles are rather weakly coupled to the rest of the system and these result in long coherence times that make them near ideal elements for a quantum computers.  Phosphorus has five valence electrons and four of these are paired up in covalent bonds with four silicon atoms, leaving the fifth unpaired.  The spins of these unpaired electrons serve as control qubits, they  couple to their atoms nuclear spin qubit as well as to neighbouring electron spin qubits.   The nuclear spins are too far apart ( $\geq$ 100 nm) for their direct dipole-dipole interactions to be significant, but they couple indirectly through the dipole-dipole interactions of their associated electron spins.

The quantum computer is then controlled by externally imposed magnetic and electric fields.  These can be varied locally at scales of around  $30 nm$,  produced by circuit elements constructed from nanowires etched onto the surface of the silicon.  These external fields move the electrons around and control the overlap of the wave functions of the electrons with the nuclear spins and with neighbouring electrons.  In this way they control exchanges of quantum information between the nuclear memory qubits and the electron control quits.

There has been significant progress towards a solid state quantum computer using phosphorus atoms doped into silicon, summarized in \cite{progress}.  But there remain challenges.

The basic idea to be explored here is that the elements of a Kane quantum computer may
also be achieved by layering a bilayer of phospholipids, which is to say a biological cell membrane or, an artificial model of a membrane, on top of a silicon wafer.  

Biological membranes are incredibly diverse, heterogeneous and complex, but their basic architecture is very 
simple\footnote{For good recent introductions, see \cite{Stillwell}.}. 
This has made it possible to construct and engineer artificial models of biological membranes which incorporate, in controlled proportions, the main elements of biological membranes.  The synthesis and study of such artificial membranes is a standard element of the tools available to biophysicists.  Also well understood, if more recent, is the layering of these artificial membranes over a silicon wafer, to enable diverse properties of them to be studied\cite{layering,talkingtochannels}.

Whether natural or artificial, the membranes are constructed from bilayers of phospholipid molecules.  Each such molecule has a ``head", which is composed of a phosphate group, possibly linked to other groups.  Two chains or tails extend downward, each composed of carbon-hydrogen units.  The heads are hydrophilic while the tails are hydrophobic so in water they arrange themselves in double layered sheets with the heads on the outer and inner surface and the tails facing into the membrane.   

Each phosphate group contains a single central phosphorus atom, covalently bonded to four oxygens. These four oxygens form a tetrahedral cage surrounding and shielding the phosphorus atom.   This leaves a fifth valence electron.  Usually this is  used up by making one of the covalent bonds to an oxygen a double bond,  which leaves no unpaired electrons on the phosphorus.  

However, we can note that some common phospholipids, such as phosphatidylcholine (PC), have an extra electron attached to their phosphate group, leading to an overall single negative charge. (See Figure 1) The extra charge is balanced by missing electrons on other components of their head groups, making these head groups polar, or zwitterionic.  
In the case of PC, the other head group is choline, composed of a nitrogen ion bonded to hydrogens.

\begin{figure}[t!]
\begin{center}
\includegraphics[width=.8 \textwidth]{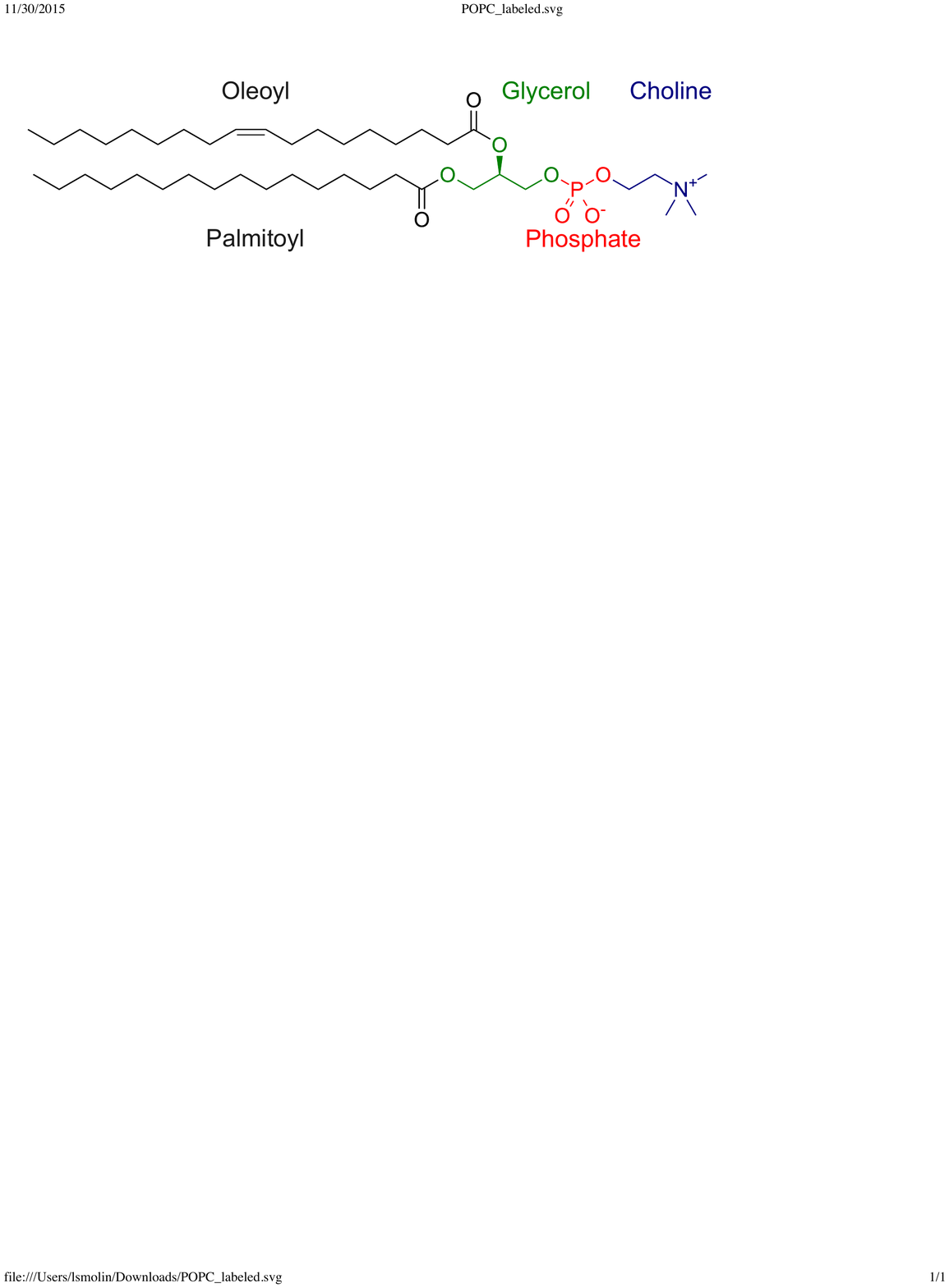}
\end{center}
\caption{The molecular structure of phosphatidylcholine (PC),  by Ed (Edgar181) - Own work. Licensed under Public Domain via Wikimedia Commons}
\label{fig1} 
\end{figure}

The extra electron on the phosphate group is unpaired and may couple through dipole dipole interactions to the nuclear spin of the phosphorus atom.  These electrons can serve as control bits, precisely analogous to the electron spins of phosphorus atoms embedded in silicon in the Kane architecture.  

The lipid composition of biological membranes is diverse and, often, heterogeneous, so the proportion of phospholipids whose head group have extra, unpaired, electrons on their phosphates varies.  In the construction of artificial, model membranes this proportion could be varied at will.   

\section{The basic concept}

The proposal is then to copy to an extent the architecture of a Kane quantum computer by layering a model, artificial, phospholipid bilayer membrane over a silicon wafer.  Some of the phosphorus nuclei in the lipid membrane play the role of memory qubits.  The control qubits are electron spins of extra electrons on charged phosphate groups in polar, zwitterionic head groups.  Their concentration can be chosen to correspond to the requirements of the quantum device.  

The concentrations of nuclear spins will be much higher than the concentration possible of classical control elements etched into the silicon.  The same may be  true of the concentration of electronic control qubits.  This means that there will be control qubits and nuclear spin qubits adjacent to any location of a classical control element.  This avoids the need to delicately align them.

On the other hand, the classical control elements, which control the control qubits through magnetic fields and electric fields, are etched on to the silicon wafer, just as in the Kane architecture.   

The original Kane proposal featured a one dimensional array of phororus atoms embedded in silicon.  There are two kinds of gates, or control elements,  built of nanowires etched onto the surface of the silicon wafer.  So called $\bf A$ gates are situated close to the phosphorus atoms.  By varying the electric charge on an $\bf A$ gate we can effect the wave function of the electron carrying the control qubit, and hence effect its coupling to the nuclear spin of that phosphorus atom.  This is used to turn on and off the coupling between the nuclear ``memory" qubit and the control electron qubit.  

The second kind of gate, $\bf J$ gates, are situated between two adjacent phosphorus atoms and control the couplings of their two electron spins by altering the overlap of their two wave functions.   

The one dimensional array is possible because the nanowires need to be spaced as closely as possible (several $nm$), so the wires run transversely to the array.   

\subsection{One dimensional arrays of qubits}

We can start by mimicking the design of a Kane quantum computer with a one dimensional array of nuclear spins.  This  consists of a set of parallel nanowires etched over the silicon chip perhaps $30 nm$ apart.  These are alternatively $A$ and $J$ gates.  By cooling the bilayer in the presence of charges on the $A$ wires, one dimensional array of $PC$ molecules is lined up perpendicularly to the direction the wires run.  This now serves as an analogue Kane quantum computer.  Charge on an $A$ gate brings the electron and nuclear spins of the adjacent $31P$ into interaction.  Charge on a $J$ gate brings two adjacent electron spins into interaction.

\subsection{Two dimensional arrays}

Architectures for two dimensional arrays are possible and involve two layers of nanowires,  whose wires run perpendicularly to each other\cite{progress}.  In devices where the phosphorus atoms are embedded in a silicon wafer, these may be etched on the top and bottom surfaces of the silicon.   In the current proposal, there may be two layers of silicon sandwiching the lipid bilayer.  

This may have several advantages over the original Kane architecture.

One difficulty implementing the original Kane proposal is getting the phosphorus atoms in the silicon arranged to correspond to the nanowire that control the exchanges and processing of quantum information.  Here we can take advantage of several properties of phospholipid membranes.   One property already mentioned is that the density in the two dimensional membrane of unpaired electrons in the phosphate groups, and hence of control qubits  carried by the electron spins, can be varied.  A second very interesting property is that the two dimensional structure of the phospholipid molecules varies according to the phase of the membrane.  Above a transition temperature, $T_c$, the molecules are in a liquid phase, in which the molecules are free to diffuse in the two dimensional membrane.  Below $T_c$ the material exists in a frozen or gel phase, where the two dimensional positions are frozen.  We can take advantage of the fact that the molecules such as $PC$ which carry extra electrons which can serve as control qubits are polar as the head groups, while neutral overall, have large electric dipole moments.  

This suggests a method of fabrication.  First etch the silicon with nanowires on which charge can be distributed as in the Kane architecture.   Lay the phospholipid membrane over the etched silicon and heat the temperature above $T_c$, so that the molecules can diffuse.  If positive and negative charges are appropriately applied to the nanowires making up the gates the more polar $PC$ molecules can be attracted to positions above the gates where we want them.  Then cool below $T_c$,  freezing them into place.  

Finally, we note that silicon is only playing a structural role, to provide a rigid frame on which to secure the arrangement of the 
$\bf A$ and $\bf J$ gates in proximity to the memory and control qubits carried by the phospholipid membrane.   

\section{Elaborations}

There are several ways this concept might be elaborated.

\subsection{Using interactions among the nuclear spins}

The phosphate groups in a phospholipid membrane are much closer, 2 to 10 angstroms apart, than are the imbedded phosphorus atoms in silicon contemplated for a solid state quantum computer.  The nuclear dipole-dipole interactions are at least  $10^6$ times stronger than the best that could be achieved in 31P doped silicon with minimum spacings of $1000$ angstroms.   So the nuclear dipole-dipole interactions may play a role in the architecture of the quantum device.

If the $31P$ nuclear spins not adjacent to control elements play no role in the architecture of the quantum computer, their dipole-dipole interactions with the memory and control qubits are a possible source of noise.   

\subsection{Making use of the bilayer for  protection of quantum information}

There is a natural mechanism for protection of quantum information due to the fact that the phospholipid membrane is a bilayer, so  we actually have two lattices of 31P nuclear spins, $50-60$ angstroms apart.   In certain phases the top and bottom head groups are lined up and well correlated  The two members of each pair experience nearly the same electric and magnetic fields coming from the silicon devices, so we may consider qubits defined by the antisymmetrization of their wave functions.

\subsection{A possible role of membrane channels}


Embedded in a biological membrane will be a variety of proteins.  Some of these serve as ion pumps, which pump specific ions across the membrane from the interior to the exterior of a cell.  This sets up a charge layer just outside the cell.  Energy is gained in a controlled way by channels-also composed of proteins, which allow the ions to flow back through the membrane back into the cell.  Some of others are coupled to the production of ATP, which are the primary means of energy storage and transmission in the cell.  But the point is that the pumps and channels are controlled by the cell and so are strongly time dependent.

For our purposes we can note that the electric currents flowing through the pumps and channels are coupled to the nuclear magnetic moments in the phosphorus atoms, through magnetic fields induced by the currents.  The electrons and nuclei are also coupled to time varying electric fields due to the buildups and discharges of the charge layers just outside the membranes.

We then may consider more ambitious designs in which control, input and readout can be achieved by modulating the ion currents flowing through the pumps and channels.  
These would make use of more of the functions 
of a biological membrane\footnote{We  may note that some of the ion channels  are voltage gated,  meaning the open in close at preset values of the electric field; others are ligand gated, which means they open and close when bound to a neurotransmitter. }.  

To achieve classical control, we may again design an interface etched into a silicon wafer over which the phospholipid membrane is layered, but this time with some of the apparatus of a  cell powering the pumps and channels.   This would either couple locally to pumps and channels in the membrane or currents could be run across the membrane by 
layering it between two silicon wafers.   

Very remarkably,  coupling between elements etched in silicon wafer and channels in a bi-lipid membrane layered over it have been achieved experimentally\cite{talkingtochannels}.

\section{Possible settings for quantum biology on neuronal membranes}

I now turn to a very brief sketch of another proposal\cite{LS-proposal} to find a role for quantum physics which uses as qubits topological degrees of freedom,
which may be induced by the chirality of the phospholipids that make up
the membrane.
  No silicon wafer is needed, and the proposal might just possibly be relevant for live as well as artificial neurons, 

\subsection{Looking for topological qubits in a neuronal membrane by ``reverse engineering"}

One way to approach the question is to assume that someone, perhaps from the future or from a secret lab, has informed us that there are quantum coherent states which arise as protected channels on the membranes of neuronal axons.  Our challenge is to reverse engineer these states.

Let's review what we have to work with, going from the outside in.

\begin{enumerate}

\item{}The cell sits in a wet medium and is largely water on the inside.

\item{}The inner  and outer surfaces of the axon membrane are sheathed in  a jacket of charged ions, mainly sodium, potassium and chlorine..  
The membrane itself is an excellent insulator, and functions like a capacitor, holding a large charge separation.
The charge densities on each side are controlled by ion gates and pumps, which are proteins, which cross the membrane.  
In a resting state, when not discharging,  these maintain a constant electric field transverse to  the membrane,
$E^{\hat{r}}_0$, 
giving rise to a voltage across the membrane of about $-70 \mu V$.  When the cell
fires, this resting potential is discharged, and then restored, in a series of events driven by opening and closing the ion channels at specific voltages, that propagate down the axon\cite{HH}.

\item{}The membrane itself is composed of  a bi-layer of phospholipid molecules, with their head groups, being hydrophilic, on the outside.  The phosphate head
groups then give us
$2d$ arrays of weakly coupled qubits, which are the $P_{31}$ nuclear spins, which couple weakly through their dipole moments.  
Each phosphate also has an unpaired electron, which gives us also a $2d$ array of electrons,
which polarize the head groups and also couple their dipole moments to each other as well as to the dipole moments of the nuclear spins.  

Phospholipids are chiral molecules, like many large biological molecules.
The tails are made of two long helical chains, that always curl in one handedness, with respect to the direction to the head. 
It has been shown that light passing through them suffers a frequency dependent rotation in their their plane of polarization, i.e. is 
birefringent\cite{Bire}.  There are also intriguing results which suggest that
the birefringence across the membrane changes in response to changes in the
potential  difference\cite{birechanges}.
Electrons traveling through chiral biological materials also are affected by the chirality\cite{Namaan}.

Electromagnetic fields interacting with a phospholipid membrane are then an example of electromagnetism coupled to chiral matter\cite{ECM}.  

\item{} The top layer of phospholipid molecules is reflected to the bottom layer, which lines up with it, tail to tail.  
So, ideally we have two lattices of potential qubits lined up against each other, which means we could seek a channel protected from thermal noise
by defining qubits as entangled singlet states of top and bottom  phosphates nuclear spins, from which noise decouples.  

Complicating this oversimplified picture is the fact that the membranes can exist in different phases; there is a frozen phase, in which we might hope to pair up top and bottom head groups, and a liquid phase, where the order is lost.  There are sometimes solid islands floating in the liquid phase.  And there are many impurities;  proteins,  chloresterals, and other molecules embeded in the membranes.

\item{} The membrane then can be regarded as a chiral insulator, made of the lipid
tails, surrounded by a conducting medium-mostly water-the intracellular and extracellular media.   The border is populated by charged ions, just under which are the magnetic and electric dipole moments of the head groups.

We note the suggestions of \cite{waveguide1,waveguide2}, who hypothesize that the axonic
membrane  functions {\it as a waveguide} for cylindrical waves, which propagate down the axon.  In fact what we have is an insulating, chiral medium, sandwiched between two conductors-the ions in the water interior and external to the
membrane.    If we want to tentatively explore the scenarios described in
\cite{waveguide1,waveguide2}, which complement  the conventional
Hodgkins-Huxley model of signal propagation\cite{HH}, we may add
the possibility that effects from the chirality of the phospholipid molecules may play a role.

\item{}  The chirality could manifest itself in several ways.  In the presence of the membrane, the chemical potentials of left and right handed fermions are different.
This could give rise to the Chiral Magnetic Effect and other similar effects\cite{ECM}.   
These would induce a surface coupling between electric and magnetic flux.
These may induce new couplings of the electric and magnetic fields to
our hypothetical qubits: the $P_{31}$ magnetic moments.  

In addition, the propagation of light through a chiral medium can induce a mass for  the photons.  This introduces a gap in the energy spectrum, below which there may reside a finite dimensional space of degenerate vacua.  These depend on the topology of the two dimensional effective geometry, and are hence called
topological degrees of freedom.  These can decouple from noise, leading to stable quantum degrees of freedom.

These topological degrees of freedom have been studied in the context of quantum materials\cite{CS-CMT} as well as extreme astrophysical environments\cite{MCS,CFJ};  to my knowledge
it has not previously been suggested that they may play a role in neuro-biology.

The worry of course is that these chiral induced couplings will be extremely
weak, and insufficient to construct a topological channel protected from thermal noise.  For this to happen the gap energy should correspond to a high temperature
compared to the ambient temperature.  Artificial membranes may be cooled to this point; whether a sufficiently high gap arises in natural biological membranes to
ensure sufficient coherence of the topological degrees of freedom is unfortunately
not  a question we can settle here.

\item{}  Within the chiral material,
the energy momentum relations of left and right handed photons are also different.  
As shown in \cite{ECM}, the different ways that chiral matter can effect the propagation of light are summarized in an effective theory 
by a one form $b_a$, with dimensions of inverse length.

In the presence of a non-vanishing
$b_a$, Maxwell's theory is altered by the addition of a four dimensional extension of Chern-Simons theory, introduced earlier by Carroll, Field,  and Jackiw\cite{CFJ}
\f
S^{CFJ} = \int d^4 x  -\frac{1}{4 e^2} \{ F_{ab} F^{ab}   
- \epsilon^{abcd} b_a A_b \partial_c A_d   \} +   J^a A_a
\label{CFJ}
\ff
The one form $b_a$ measures the chirality properties of the membrane.
$b_0$ is proportional to the chiral chemical potential, $\Delta \mu$, which is defined as the
difference between the chemical potentials of left and right handed particles\cite{ECM},
\f
b_0 = \frac{e^2}{4 \hbar} \Delta \mu
\ff
$\Delta \mu$ is not equal to zero when it takes more or less energy for the left handed polarization of light to travel on the helical chains of the tails of the phospholipid, than it does the right handed.                                                                        

We see that gauge invariance of ( \ref{CFJ}  ) imposes.
\f
db=0
\ff
which implies that locally $b= d\phi $ for some scalar field $\phi$.

The first pair of Maxwell equations remains unaltered as it expresses that
$F= dA$.

The second set of Maxwell equations are altered.
\f
\partial_a F^{ab}= e^2 J^b - \epsilon^{bcde} b_c \partial_d A_e
\label{MCS1}
\ff

These become

\f
\nabla \times B^i -   \frac{\partial E^i }{dt} = e^2 J^i + b_0 B^i +  ( b \times E )^i
\label{MCS1A}
\ff
\f
\nabla \cdot E=  e^2 J^0  -  b \cdot B
\label{MCS1B}
\ff

We can then decompose these into components in $\hat{r}, \hat{\alpha} = 1,2   $

\f
\partial_{[\alpha}   B_{\beta]} -   \frac{\partial E^z }{dt} = e^2 J^z + b_0 B^z +   ( b_{[\alpha } E_{\beta]} )
\label{MM1}
\ff
\f
\epsilon^{\alpha \beta }\partial_{[\beta}   B_{z]} -   \frac{\partial E^\alpha }{dt} 
= e^2 J^\alpha + b_0 B^\alpha  +   ( b_{[\alpha } E_{\beta]} )
\label{MM2}
\ff
\f
\nabla \cdot E= e^2 J^0  -  b \cdot B
\label{MM3}
\ff
\f
\nabla^z E_z + \nabla^\alpha E_\alpha = e^2 J^0  -  b^z  B_z + b^\alpha  B_\alpha 
\label{MM4}
\ff

One consequence is that for either or both $b_0 \neq $ and $b_i \neq 0$,
one or both polarizations of the photon are gapped when traveling in the
chiral medium.     


We may then model the chiral physics of the membrane of an axon as a long cylinder, where
the longitudinal coordinate is $\hat{x}$, the radial coordinate is $\hat{r}$
and the angular coordinate on the $S^1$ is $\phi$.  We will
sometimes write $\alpha, \beta = 1,2$, for general coordinates on the cylinders
of constant $r$.     

The chirality vector, $b_i$ is along $\hat{r}$ and vanishes outside of the thickened
cylinder,   $a \geq r \geq   b$. It points in the opposite way in the two layers, as they are tail to tail.  

\item{}The gap and the mass of the photons is proportional to the magnitude of $b_0$.  At very low temperatures, ${\cal T} <<  b_0$  we can decouple the local degrees of freedom, and we are left with an abelian topological
field theory governing the low energy physics.  This has a representation
given by giving amplitudes to a set of non-crossing holonomies of abelian, $U[1]$
gauge theory on a punctured cylinder, where the punctures represent the ion channels which cross the membrane.

A basis for the Hilbert space of the topological sector is then given by holonomies on the $\alpha$
loops, which circle the cylinder, ,separated by punctures.   
These are defined as
\f
T[\alpha , A] = e^{\imath \oint_\alpha A }
\label{holonomy}
\ff

We can write this as a Fock space built of functionals
\f
\Psi (A) = f( T[ \alpha_1, A ] , T[ \alpha_2, A ], \dots   ) 
\ff
A basis is given by functionals of the form 
\f
T[\{ w_i , \}] =  T[ \alpha_i^{w_i}, A ] = T[ \alpha_i, A ]^{w_i} 
\ff
which are the windings of the cylinder, with winding number given by $w_i$.  

\item{}When we turn the temperature on,   we couple the massive states to the topological states, which will introduce thermal noise at scales above 
the gap $|b| = L^{-1} $.    A very rough order of magnitude estimate, in which
we take $L$ to be the distance the helical tail takes to wind once, suggests the gap
could be above room temperature, making the topological winding states at least metastable over signal propagation times.

\item{} A key question is whether there are further strategies to
decouple at least some of the topological states from noise.   

For example, can we
expect that the random noise spectrum is invariant under
parity?  This could means the noise only couples to those states
$T[\{ w_i \ = \  \mbox{even } \}  ]$.     This leaves a protected channel consisting of
states with odd windings.

Might it also be possible to decouple winding states from the thermal noise if they wind around a large number of punctures, so that they loop a long distance along the axon?  

The question we want to ask is if, in the large number of winding states that will be found on or in the membrane of a long axon, are there any that sufficiently
decouple from the thermal noise as to give us a noise-free subsystem,
given the values of $b_a$  caused by the thin layers of chiral molecules?

At this first discussion we can make no claims.


\item{} But let's go a few more steps with the reverse engineering.  Lets look at the first modified Maxwell equation (\ref{MM1}).   Focus on the transverse magnetic
and electric fields, $E^r$ and $B^r$.
\f
 -   \frac{\partial E^r }{dt} = J^r + b_0 B^r 
 \label{MM10}
\ff
This gives first of all the proportionate relation between $\dot{E}^r$ and
the current $J^r$, but the current flow now powers also a transverse magnetic
field through the last term.  This is relate.  This may be represented on loop
states as,
\f
\frac{{\cal E}[R]}{dt}- {\cal J}[R] = b_0 \int_R B^{r} \approx b_0 [T[\beta, A ]  -1]
\ff
where $\beta$ is a loop that surrounds the puncture, bounding an area $R$,
while ${\cal E}[R] $ and ${\cal J}[R] $ are the flux of elecytric field and current
through that area.
This can then act to transfer a topological loop quantum across  the 
membrane channel.

 This may also couple to  the arrays of nuclear spins
of the phosphates.  That suggests that when the neuron is discharging, releasing
the voltage by letting the ions flow back through the membrane channels to restore equilibrium, the result may be a perturbation or measurement of the 
qubits protected in the phosphate head groups.

\end{enumerate}


\section{Conclusions}

In this short  communication we have sketched two ideas for what can be called ``bionic" quantum devices.  The first employs a hybrid of silicon wafer technology and model biological membranes.  The second uses the natural  membrane 
of an axon as a wave guide, following \cite{waveguide1,waveguide2},  
a role it may or may not play in nature.
The second also employs the chirality of the phospholipid components of
the membrane to induce Chern-Simons couplings, opening the door to abelian
topological winding states.

 There are many issues that must be confronted before it can be argued that either of these are workable ideas.

We mention some key open issues, first with respect to the bionic-Kane architecture:

\begin{itemize}

\item{}Does the hybrid device need to be cooled for coherence times to be long enough?

\item{}The phosphorus atoms make a two dimensional array in the membrane, can the control elements etched into silicon be two dimensional.  

\end{itemize}

Finally, it has not of course escaped our attention that if you take away the silicon wafer you still may in some conditions have a device that  can store and process quantum information in a way that interacts with the classical degrees of freedom that control the pumps and channels.

The key question regarding the idea that topological excitations of the electromagnetic field are induced by passage through the chiral membrane is whether the gap is large enough to decouple the topological modes.

Whether real biological cells make use of this opportunity is a much harder question.   
In the preceding section, we put forward a very tentative suggestion as to how this might come about.  

\section*{ACKNOWLEDGEMENTS}

I am grateful to Amy Arnstein, David Cory, Kamran Diba, Richard Epand,  Raymond Laflamme, Jaron Lanier,   Roger Melco,   Michele Mosca,  Ron Naaman,  Gil Prive,  
Rob Spekkens and Chong Wang for conversations about aspects of these ideas.
  Maikel C. Rheinst\"{a}dter and Christoff Simon, also provided very useful comments on drafts of this paper.    

This research was supported in part by Perimeter Institute for Theoretical Physics. Research at Perimeter Institute is supported by the Government of Canada through Industry Canada and by the Province of Ontario through the Ministry of Research and Innovation. This research was also partly supported by grants from NSERC and FQXi.  I am especially thankful to the John Templeton Foundation for their generous support of this project.


\end{document}